\documentclass[twocolumn,british, superscriptaddress]{revtex4-1}
\usepackage[T1]{fontenc}
\usepackage[latin9]{inputenc}
\usepackage{amsmath}
\usepackage{amssymb}
\usepackage{graphicx}
\usepackage{babel}

\usepackage{color}

\begin{document}

\title{Accurate Interatomic Force Fields via Machine Learning with Covariant Kernels}

\author{Aldo Glielmo}
\email{aldo.glielmo@kcl.ac.uk}
\affiliation{Department of Physics, King's College London, Strand, London WC2R 2LS, United Kingdom}
\author{Peter Sollich} 
\affiliation{Department of Mathematics, King's College London, Strand, London WC2R 2LS, United Kingdom}
\author{Alessandro De Vita}
\affiliation{Department of Physics, King's College London, Strand, London
  WC2R 2LS, United Kingdom} 
\affiliation{Dipartimento di Ingegneria e
  Architettura, Università di Trieste, via A. Valerio 2, I-34127 Trieste, Italy}

\begin{abstract}
	
We present a novel scheme to accurately predict atomic forces as vector
quantities, rather than sets of scalar components, by Gaussian Process
(GP) Regression. This is based on matrix-valued kernel functions, on
which we impose the requirements that the predicted force rotates with the target configuration
and is independent of any rotations applied to the configuration database
entries. We show that such covariant GP kernels can be obtained
by integration over the elements of the rotation group $SO(d)$ for
the relevant dimensionality $d$. 
Remarkably, in specific cases the
integration can be carried out analytically and yields a conservative
force field that can be recast into a pair interaction form. 
Finally,
we show that restricting the integration to a summation over the elements
of a finite point group relevant to the target system is sufficient
to recover an accurate GP. The accuracy of our kernels in predicting
quantum-mechanical forces in real materials is investigated by tests
on pure and defective Ni, Fe and Si crystalline systems.
\end{abstract}

\maketitle

\section{Introduction }

Recent decades have witnessed an exponential growth of computer
processing power (\textquotedblleft Moore's Law\textquotedblright{}
\cite{Moore:1998gt}) and an equally fast progress of storage technology
(\textquotedblleft Kryder\textquoteright s Law\textquotedblright{}
\cite{Walter:2005be,Grochowski:2003gg}). Atomistic modelling methods
based on computation and data-intensive quantum mechanical methods,
such as Density Functional Theory (DFT) \cite{Hartree:2008fd,Fock:1930dx,Slater:1951en},
have correspondingly evolved in both feasibility and scope. Moreover,
the possibility of retaining at low cost very large amounts of data
generated by Quantum Mechanical (QM) codes has prompted novel efforts
to make the data openly accessible \cite{Untitled:Bxh2cefs}. 

The information contained in the data can thus be harnessed and re-used
indefinitely, in various ways. High throughput techniques are routinely
used to identify new correlations between physical properties, with
the aim of designing new high-performance materials \cite{Ghiringhelli:2015kr,Pilania:2013di,Kim:2016ev}.
Inference techniques can meanwhile also be used as a boost or substitute
for QM techniques. This typically involves predicting a physical property
for a new system configuration, on the basis of its values for an
existing database of configurations. If the database is sufficiently
large and representative, the new property values can be quickly inferred,
rather than calculated anew by expensive QM procedures, with controllable
accuracy. 

Machine Learning techniques have been successfully used to predict
properties as diverse as atomisation energies \cite{Rupp:2012kxa},
density functionals \cite{Snyder:2012da}, Green's functions \cite{Arsenault:2014ema},
electronic transport coefficients \cite{LopezBezanilla:2014iq}, potential
energy surfaces \cite{Bartok:2010fj,Behler:2007fe,Shapeev:2016kn} and free energy
landscapes \cite{Stecher:2014gi}. The high configuration space complexity
of real chemical systems has also inspired ``learning'' molecular dynamics
schemes that never assume database completeness, but rather combine
inference with on-the-fly QM calculations (learning on-the-fly, LOTF)
\cite{DeVita:1997fp,Csanyi:2004dh,Podryabinkin:2016ta} carried out when inference is
infeasible or not deemed sufficiently accurate. 

A well established general concept within the Machine Learning community
is that functional invariance properties under some known transformation
can be used to improve prediction, whether this is carried out by e.g.\
Gaussian Process (GP) regression \cite{Haasdonk:2007ff,Williams:2006vz}
or neural networks \cite{Bishop:998831}. 
Exploiting in similar ways
properties other than invariance has received more limited attention
\cite{Krejnik:2012eo}. In the same spirit, materials modellers have
been successful in exploiting the invariance of energy under rotation
or translation to improve the performance of energy prediction techniques
\cite{Bartok:2010fj,Behler:2007fe}.
In LOTF molecular dynamics applications
the high-accuracy target and local interpolation character of force
prediction makes it appealing to learn forces directly rather than
learning a potential energy scalar field first and then deriving forces
by differentiation. In previous works \cite{Li:2015eb,Caccin:2015co,Botu:2014kc}
this was accomplished by using GP regression to separately learn individual
force components.

Here, we show how \emph{Vectorial} Gaussian Process (VGP) \cite{Micchelli:2004uv,Alvarez:2012ew} regression provides a more natural
framework for force learning, where the correct vector behaviour of
forces under symmetry transformations can be obtained by using a new
family of vector kernels of \emph{covariant} nature. These kernels
prove particularly efficient at exploiting the information contained
in QM force databases, however constructed, together with any prior
knowledge of the symmetry properties of the physical system under
investigation. The next section provides a brief overview of the notion
of a VGP, where we pay particular attention
to the problem of force learning. Then we define a covariant kernel,
explain its symmetry properties and give a general recipe to generate
such kernels. The procedure is best exemplified by looking at one
and two dimensional systems, where the relevant symmetry force transformation
groups are $D_{1}$ and $O(2)$. Finally we address the full three
dimensional case, where covariant kernels are tested by examining
their performance in learning QM forces in realistic physical systems.

\section{Vectorial Gaussian Process Regression}

We wish to model by a VGP the force $\mathbf{f}$ acting on an atom
whose chemical environment is in a configuration $\mathbf{\rho}$
that encodes the positions of all neighbours of the atom, up to
a suitable cutoff radius, in an arbitrary Cartesian reference frame.
In the absence of long range ionic interactions, the existence of
  such a local map is guaranteed for all finite-temperature systems by the "nearsightedness" principle of electronic matter \cite{Kohn:1996hn,Prodan:2005eu}.

In a Bayesian setting, before any data is considered, $\mathbf{\mathbf{f}}$
is treated as a Gaussian Process, i.e., it is assumed that for any
finite set of configurations $\{\rho_{i,\,}i=1,\dots N\}$ the values
$\mathbf{f}(\rho_{i})$ taken by the vector function $\mathbf{f}$
are well described by a multivariate Gaussian distribution \cite{Williams:2006vz}.
We write: 
\begin{equation}
\mathbf{f}(\rho)\sim\mathcal{GP}(\mathbf{m}(\rho),\mathbf{K}(\rho,\rho'))\label{eq: prior gp}
\end{equation}
where $\mathbf{m}(\rho)$ is a vector-valued mean function and $\mathbf{K}(\rho,\rho')$
is a matrix-valued kernel function. Before any data is considered,
$\mathbf{m}$ is usually assumed to be zero as all prior information
on $\mathbf{f}$ is encoded into the kernel function $\mathbf{K}(\rho,\rho')$.
The latter represents the correlation of the vectors $\mathbf{f}(\rho)$
and $\mathbf{f}(\rho')$ as a function of the two configurations (``input
space points'') $\rho$ and $\rho'$: 
\begin{equation}
\mathbf{K}(\rho,\rho')=\langle\mathbf{f}(\rho)\mathbf{f}^{{\rm T}}(\rho')\rangle,\label{eq: prior kernel}
\end{equation}
where angular brackets here signify the expected value over the multivariate
Gaussian distribution. Any kernel $\mathbf{K}$ consistent with this
definition must be a positive semi-definite matrix function, since
for any collection of vectors $\{\mathbf{v}_{i}\}$ 
\begin{equation}
\sum_{ij}\mathbf{v}_{i}^{{\rm T}}\mathbf{K}(\rho_{i},\rho_{j})\mathbf{v}_{j}=\langle(\sum_{i}\mathbf{v}_{i}^{{\rm T}}\mathbf{f}(\rho_{i}))^{2}\rangle\ge0.\label{eq: PSD proof}
\end{equation}

To train the prediction model we need to access a database of atomic
configurations and reference forces $\mathcal{D}=\{(\rho,\mathbf{f}^{r})_{i},\,i=1,\dots,N\}$.
Using Bayes\textquoteright{} theorem \cite{Bayes:1763ug} the distribution
(\ref{eq: prior gp}) is modified to take the data $\mathcal{D}$
into account \cite{Williams:2006vz}. If the \emph{likelihood function}
\cite{Bishop:998831} is also Gaussian (which effectively assumes
that the observed forces $\ensuremath{\mathbf{f}_{i}^{r}}$ are the
true forces subject to Gaussian noise of variance $\sigma_{n}^{2}$)
then the resulting \emph{posterior distribution} $\mathbf{f}(\rho\mid\mathcal{D})$,
conditional on the data, will also be a Gaussian process 
\begin{equation}
\mathbf{f}(\rho\mid\mathcal{D})\sim\mathcal{GP}(\mathbf{\hat{f}}(\rho\mid\mathcal{D}),\mathbf{\hat{C}}(\rho,\rho')).\label{eq: posterior gp}
\end{equation}
The mean function of the posterior distribution, $\hat{\mathbf{f}}(\rho\mid\mathcal{D})$,
is at this point the best estimate for the true underlying function:
\begin{equation}
\hat{\mathbf{f}}(\rho\mid\mathcal{D})=\sum_{ij}^{N}\mathbf{K}(\rho,\rho_{i})[\mathbb{K}+\mathbb{I}\sigma_{n}^{2}]_{ij}^{-1}\mathbf{f}_{j}^{r}.\label{eq: posterior mean}
\end{equation}
Here $\sigma_{n}^{2}$, formally the noise affecting the observed
forces $\mathbf{f}^{r}$, serves in practice as a regulariser for
the matrix inverse. In the following, blackboard bold characters such
as $\mathbb{K}$ or $\mathbb{I}$ indicate $N\times N$ block matrices
(for instance, the \emph{Gram matrix} $\mathbb{K}$ is defined as
$(\mathbb{K})_{ij}=\mathbf{{K}}(\rho_{i},\rho_{j})$). Similarly,
we denote by $[\mathbb{K}+\mathbb{I}\sigma_{n}^{2}]_{ij}^{-1}$ the
$ij$-block of the inverse matrix.

We next examine how to incorporate the vector behaviour of forces
into the learning algorithm. The relevant symmetry transformations
in the input space are: rigid translation of all atoms, permutation
of atoms of the same chemical species, rotations and reflections of
atomic configurations. Forces are invariant with respect to translations
and atomic permutations, and covariant with respect to rotations and
reflections. Assuming that the representation of the atomic configuration
is local, i.e., the atom subject to the force $\mathbf{f}_{i}$ is
at the origin of the reference frame used for $\rho_{i}$, translations
are automatically taken into account. The remaining symmetries must
be addressed in the construction of covariant kernels. 

\section{Covariant Kernels}

From now on we will define $\mathcal{S}$ to be any symmetry operator
(rotation or reflection) acting on an atomistic configuration of a
$d$-dimensional system. Rotations will be denoted by $\mathcal{R}$
and reflections by $\mathcal{Q}$.

We require two properties to apply to the predicted force $\hat{\mathbf{f}}(\rho\mid\mathcal{D})$,
once configurations are transformed by an operator $\mathcal{S}$
(represented by a matrix $\mathbf{S}$):

\paragraph*{Property 1}

If the target configuration $\rho$ is transformed to $\mathcal{S}\rho$,
the predicted force must transform accordingly:
\begin{equation}
\hat{\mathbf{f}}(\mathcal{S}\rho\mid\mathcal{D})=\mathbf{S}\hat{\mathbf{f}}(\rho\mid\mathcal{D}).\label{eq: property 1}
\end{equation}

\paragraph*{Property 2}

The predicted force must not change if we arbitrarily transform the
configurations in the database ($\mathcal{D}\rightarrow\tilde{\mathcal{D}}=\{(\mathcal{S}_{i}\rho_{i},\mathbf{S}_{i}\mathbf{f}_{i}^{r})\}$)
with any chosen set of roto-reflections $\{\mathcal{S}_{i}\}$.

We next introduce a special class of kernel functions that automatically
guarantees these two properties: a \emph{covariant kenrel} has the
defining property
\begin{equation}
\mathbf{K}(\mathcal{S}\rho,\mathcal{S}'\rho')=\mathbf{S}\mathbf{K}(\rho,\rho')\mathbf{S}'^{{\rm T}}.\label{eq: covariant property}
\end{equation}
That a covariant kernel imposes Property 1 follows straightforwardly
from Eq.~(\ref{eq: posterior mean}):
\begin{eqnarray}
\hat{\mathbf{f}}(\mathcal{S}\rho\mid\mathcal{D}) & = & \sum_{ij}^{N}\mathbf{K}(\mathcal{S}\rho,\rho_{i})[\mathbb{K}+\mathbb{I}\sigma_{n}^{2}]_{ij}^{-1}\mathbf{f}_{j}^{r}\nonumber \\
 & = & \sum_{ij}^{N}\mathbf{S}\mathbf{K}(\rho,\rho_{i})[\mathbb{K}+\mathbb{I}\sigma_{n}^{2}]_{ij}^{-1}\mathbf{f}_{j}^{r}\nonumber \\
 & = & \mathbf{S}\hat{\mathbf{f}}(\rho\mid\mathcal{D}).\label{eq: Prop 1 Proof}
\end{eqnarray}
To prove Property 2 we note that, if the kernel function is covariant,
the transformed database $\tilde{\mathcal{D}}$ has Gram matrix $(\mathbb{\tilde{K}})_{ij}=\mathbf{K}(\mathcal{S}_{i}\rho_{i},\mathcal{S}_{j}\rho_{j})=\mathbf{S}_{i}\mathbf{K}(\rho_{i},\rho_{j})\mathbf{S}_{j}^{{\rm T}}$.
If we define the block-diagonal matrix $\mathbb{S}_{ij}=\delta_{ij}\mathbf{S}_{i}$,
this can be written in the simple block-matrix form $\mathbb{\tilde{K}}=\mathbb{\mathbb{S}K}\mathbb{S}^{{\rm T}}$.
Using kernel covariance again to write $\mathbf{K}(\rho,\mathcal{S}_{i}\rho_{i})=\mathbf{K}(\rho,\rho_{i})\mathbb{S}_{ii}^{{\rm T}}$
the prediction associated with the transformed database $\tilde{\mathcal{D}}$
can be written as 
\begin{equation}
\hat{\mathbf{f}}(\rho\mid\tilde{\mathcal{D}})=\sum_{ij}^{N}\mathbf{K}(\rho,\rho_{i})\mathbb{S}_{ii}^{{\rm T}}[\mathbb{S}\mathbb{K}\mathbb{S}^{{\rm T}}+\mathbb{I}\sigma_{n}^{2}]_{ij}^{-1}\mathbb{S}_{jj}\mathbf{f}_{j}^{r}.\label{eq: transformed database prediction}
\end{equation}
By simple matrix manipulations it is now possible to show that in
the above expression the symmetry transformations cancel out; indeed
\begin{eqnarray}
\mathbb{S}^{{\rm T}}[\mathbb{S}\mathbb{K}\mathbb{S}^{{\rm T}}+\mathbb{I}\sigma_{n}^{2}]^{-1}\mathbb{S} & = & \mathbb{S}^{{\rm T}}[\mathbb{S}(\mathbb{K}+\mathbb{I}\sigma_{n}^{2})\mathbb{S}^{{\rm T}}]^{-1}\mathbb{S}\nonumber \\
 & = & \mathbb{S}^{{\rm T}}(\mathbb{S}^{{\rm T}})^{-1}[\mathbb{K}+\mathbb{I}\sigma_{n}^{2}]^{-1}\mathbb{S}^{{\rm -1}}\mathbb{S}\nonumber \\
 & = & [\mathbb{K}+\mathbb{I}\sigma_{n}^{2}]^{-1}\label{eq: Gram invariance}
\end{eqnarray}
Equation (\ref{eq: Gram invariance}) along with (\ref{eq: transformed database prediction})
implies $\hat{\mathbf{f}}(\rho\mid\mathcal{\tilde{{D}}})$ = $\hat{\mathbf{f}}(\rho\mid\mathcal{D})$,
that is, Property 2. It is easy to check that standard kernels such
as the \emph{squared exponential} \cite{Bishop:998831} or the overlap
integral of atomic configuration \cite{Ferre:2015dq} do not possess
the covariance property (\ref{eq: covariant property}). Designing,
entirely by feature engineering, a covariant kernel is in principle
possible but can require complex tuning and is likely to be highly
system dependent (see e.g. \cite{Li:2015eb}). We note that 
non covariant kernels can be used and avoid these difficulties, 
and some have been successfully implemented \cite{Botu:2014kc,Botu:2015kb}. 
This leaves space for improvement as prediction efficiency  will
generally be enhanced by increased exploitation of symmetry (see e.g.,
Figure 3 below for a simple test of this). 

We next present a general method for transforming a standard matrix kernel into a covariant one, followed by numerical tests suggesting that the resulting
kernel improves very significantly on the force-learning properties
of the initial one, its error converging with just a fraction of the
training data. This proceeds along the lines of previous techniques for generating scalar invariants, namely the transformation
integration procedure developed in \cite{Haasdonk:2007ff} and the 
Smooth Overlap of Atomic Orbitals (SOAP) representation for learning potential energy surfaces of atomic systems \cite{Bartok:2013cs,Bartok:2015iw}.

Given a group $\mathcal{S}$ and a \emph{base kernel} $\mathbf{K}^{b}$, a covariant kernel $\mathbf{K}^{c}$ can be constructed by
\begin{equation}
\mathbf{K}^{c}(\rho,\rho')=\int d\mathcal{S}_{1}d\mathcal{S}_{2}\mathbf{\,S}_{1}^{{\rm T}}\mathbf{K}^{b}(\mathcal{S}_{1}\rho,\mathcal{S}_{2}\rho')\mathbf{S}_{2}\label{eq: covariantdouble integration}
\end{equation}
where $d\mathcal{S}$ is the normalised Haar measure for the symmetry
group we are integrating over \cite{Mehta:1086514}.

The covariance of $\mathbf{K}^{c}$ as given by (\ref{eq: covariantdouble integration})
is easily checked as
\begin{eqnarray}
\mathbf{K}^{c}(\mathcal{S}\rho,\mathcal{S}'\rho') & = & \int d\mathcal{S}_{1}d\mathcal{S}_{2}\mathbf{\,S}_{1}^{{\rm T}}\mathbf{K}^{b}(\mathcal{S}_{1}\mathcal{S}\rho,\mathcal{S}_{2}\mathcal{S}'\rho')\mathbf{S}_{2}\nonumber \\
 & = & \int d\tilde{\mathcal{S}}_{1}d\tilde{\mathcal{S}}_{2}\mathbf{\,S}\tilde{\mathbf{S}}_{1}^{{\rm T}}\mathbf{K}^{b}(\mathcal{\tilde{S}}_{1}\rho,\tilde{\mathcal{S}}_{2}\rho')\tilde{\mathbf{S}}_{2}\mathbf{S}'^{{\rm T}}\nonumber \\
 & = & \mathbf{S}\mathbf{K}^{c}(\rho,\rho')\mathbf{S}'^{{\rm T}}\label{eq: proof of covariance}
\end{eqnarray}
where the second line follows from the substitutions $\tilde{\mathcal{S}}_{1}=\mathcal{S}_{1}\mathcal{S}$
and $\mathcal{\tilde{S}}_{2}=\mathcal{S}_{2}\mathcal{S}'$. Note that
these transformations have unit Jacobian because of the translational
invariance (within the group) of any Haar measure \cite{Aubert:2003hs,Mehta:1086514}.

It can be shown that the positive semi-definiteness of the base kernel
is preserved under the operation (\ref{eq: covariantdouble integration})
of covariant integration. In particular, a kernel is positive semi-definite
if and only if it is a scalar product in some (possibly infinite dimensional)
vector space \cite{Mercer:1909wb,Williams:2006vz}. Hence the base
kernel can be written as $\mathbf{K}^{b}(\rho,\rho')=\int d\alpha\boldsymbol{\,\phi}_{\alpha}(\rho)\boldsymbol{\phi}_{\alpha}^{{\rm T}}(\rho').\label{eq: base kernel expansion}$
It is then possible to show that its covariant counterpart $\mathbf{K}^{c}$
[Eq. (\ref{eq: covariantdouble integration})] will also be a
scalar product in a new function space. Indeed
\begin{eqnarray}
\mathbf{K}^{c}(\rho,\rho') & = & \int d\mathcal{S}_{1}d\mathcal{S}_{2}\mathbf{\,S}_{1}^{{\rm T}}\mbox{\ensuremath{\mathbf{K}}}^{b}(\mathcal{S}_{1}\rho,\mathcal{S}_{2}\rho')\mathbf{S}_{2}\nonumber \\
 & = & \int d\alpha\,d\mathcal{S}_{1}d\mathcal{S}_{2}\,\mathbf{S}_{1}^{{\rm T}}\boldsymbol{\phi}_{\alpha}(\mathcal{S}_{1}\rho)\boldsymbol{\phi}_{\alpha}^{{\rm T}}(\mathcal{S}_{2}\rho')\mathbf{S}_{2}\nonumber \\
 & = & \int d\alpha\boldsymbol{\,\psi}_{\alpha}(\rho)\boldsymbol{\psi}_{\alpha}^{{\rm T}}(\rho')\label{eq: proof of PD}
\end{eqnarray}
where the new basis vectors were defined as
$ \boldsymbol{\psi}_{\alpha}(\rho)=\int d\mathcal{S}\,\mathbf{S}^{{\rm T}}\boldsymbol{\phi}_{\alpha}(\mathcal{S}\rho).\label{eq: covariant kernel basis}$
Hence, $\mathbf{K}^{c}$ will also be positive definite.

The completely general procedure above can be cumbersome to apply
in practice, because of the double integration over group elements
in (\ref{eq: covariantdouble integration}) and the dependence
on the design of the base kernel matrix $\mathbf{K}^{b}$. As a simplification,
we assume the base kernel to be of diagonal form; assuming equivalence
of all space directions, we can then write
\begin{equation}
\mathbf{K}^{b}(\rho,\rho')=\mathbf{\mathbf{I}}k^{b}(\rho,\rho').\label{eq: isotropic base kernel}
\end{equation}
where the scalar base kernel $k^{b}$ is independent on the reference
frame in which the configurations are expressed. This requires that
\begin{equation}
k^{b}(\mathcal{S}\rho,\mathcal{S}\rho')=k^{b}(\rho,\rho'),\label{eq: global invariance}
\end{equation}
that is, scalar invariance of the base kernel (a property very commonly
found in standard kernels). The double integration in (\ref{eq: covariantdouble integration})
reduces at this point to a single one
\begin{eqnarray}
\mathbf{K}^{c}(\rho,\rho') & = & \int d\mathcal{S}_{1}d\mathcal{S}_{2}\mathbf{\,S}_{1}^{{\rm T}}\mathbf{S}_{2}k^{b}(\mathcal{S}_{1}\rho,\mathcal{S}_{2}\rho')\nonumber \\
 & = & \int d\mathcal{S}_{1}d\mathcal{S}_{2}\,\mathbf{S}_{1}^{{\rm T}}\mathbf{S}_{2}k^{b}(\rho,\mathcal{S}_{1}^{{\rm -1}}\mathcal{S}_{2}\rho')\nonumber \\
 & = & \int d\mathcal{S}\,\mathbf{S}\,k^{b}(\rho,\mathcal{S}\rho')\label{eq: covariant integration}
\end{eqnarray}
where the second line follows from property (\ref{eq: global invariance})
and the third line is obtained by the substitution $\mathcal{S}=\mathcal{S}_{1}^{{\rm -1}}\mathcal{S}_{2}$. 

In the next section we show that some base kernels allow analytical
integration of (\ref{eq: covariant integration}). Here we note that
incorporating our prior knowledge of the correct behaviour of forces
in the kernel enables us to learn and predict forces associated with
any configuration, regardless of its orientation. However, being
able to do this for completely generic orientations is not always
necessary. In many systems (e.g. crystalline solids where the orientation
is known) all relevant configurations cluster around particular discrete
symmetries. For these systems the relevant physics can be captured
by restricting Eq. (\ref{eq: covariantdouble integration}) to
a discrete sum over the relevant group elements:
\begin{equation}
\mathbf{K}^{c}(\rho,\rho')=\frac{1}{|G|}\sum_{\mathcal{G}\in G}\mathbf{G}k(\rho,\,\mathcal{G}\rho'),\label{eq: covariant summation}
\end{equation}
and since there are 48 distinct group elements at most (the order
of the full $O_{48}$ group), the procedure remains computationally
feasible. In the particular case of one-dimensional systems, where
the only available symmetry operation other than the identity is the
inversion, Eqs. (\ref{eq: covariant integration}) and (\ref{eq: covariant summation})
are formally equivalent. 

\section{Covariant Kernels from 1 to 3 dimensions}

In the following we will assume that a single chemical species is
present, so that permutation invariance will be simply enforced by
representing configurations as linear combinations of $n$ Gaussian
functions each centred on one atom, all having the same width $\sigma$,
and suitably normalised depending on the dimension $d$ considered: 
\begin{equation}
\rho(\mathbf{r},\{\mathbf{r}_{i}\})=\frac{1}{(2\pi\sigma^{2})^{d/2}}\sum_{i}^{n}\mathrm{e}^{-\frac{\|\mathbf{r}-\mathbf{r}_{i}\|^{2}}{2\sigma^{2}}}.\label{eq: configuration expansion}
\end{equation}
 From (\ref{eq: configuration expansion}), a linear base kernel $k_{L}^{b}$
can be defined as the overlap integral of two configurations \cite{Ferre:2015dq,Bartok:2010fj}
\begin{eqnarray}
k_{L}^{b}(\rho,\rho') & = & \int d\mathbf{r}\,\rho(\mathbf{r},\{\mathbf{r}_{i}\})\rho'(\mathbf{r},\{\mathbf{r}'_{j}\})\nonumber \\
 & = & \frac{1}{(2\pi\sigma^{2})^{d}}\sum_{ij}^{nn'}\int d\mathbf{r}\,\mathrm{e}^{-\frac{\|\mathbf{r}-\mathbf{r}_{i}\|^{2}}{2\sigma^{2}}}\mathrm{e}^{-\frac{\|\mathbf{r}-\mathbf{r}'_{j}\|^{2}}{2\sigma^{2}}}\nonumber \\
 & = & \frac{1}{(2\sqrt{\pi\sigma^{2}})^{d}}\sum_{ij}\mathrm{e}^{-\frac{\|\mathbf{r}_{i}-\mathbf{r}'_{j}\|^{2}}{4\sigma^{2}}}\label{eq: base overalp kernel}
\end{eqnarray}
where the integration yielding the third line is performed by standard
completion of the square.

We can interpret the linear kernel $k_{L}^{b}$ in (\ref{eq: base overalp kernel})
as a scalar product in function space, so that $k_{L}^{b}(\rho,\rho)=\|\rho\|^{2}$
can be thought of as the squared norm of the $\rho$ configuration
function. A permutation invariant distance is also readily obtained
as $d(\rho,\rho')=\|\rho-\rho'\|$, which can be used within a squared
exponential kernel to give 
\begin{eqnarray}
k_{SE}^{b}(\rho,\rho') & = & \mathrm{e}^{-\|\rho-\rho'\|^{2}/2\theta}\nonumber \\
 & = & \mathrm{e}^{-(k_{L}^{b}(\rho,\rho)+k_{L}^{b}(\rho',\rho')-2k_{L}^{b}(\rho,\rho'))/2\theta}.\label{eq: SE kernel}
\end{eqnarray}

The representation described above is by construction translation (and atomic
permutation) invariant. We next address the transformations for which
the atomic force is covariant, i.e., rotations and reflections, using
the approach described in the previous section. Systems with dimensions
$d=1,2,3$ are considered in the following three subsections. The
first two provide a useful conceptual playground where the features
of ``covariant learning'' can be more easily visualised. The third
one benchmarks the method in real physical systems, simulated at the
DFT level of accuracy.

\subsection{1D systems}

\begin{figure}
\begin{centering}
\includegraphics[width=0.8\columnwidth]{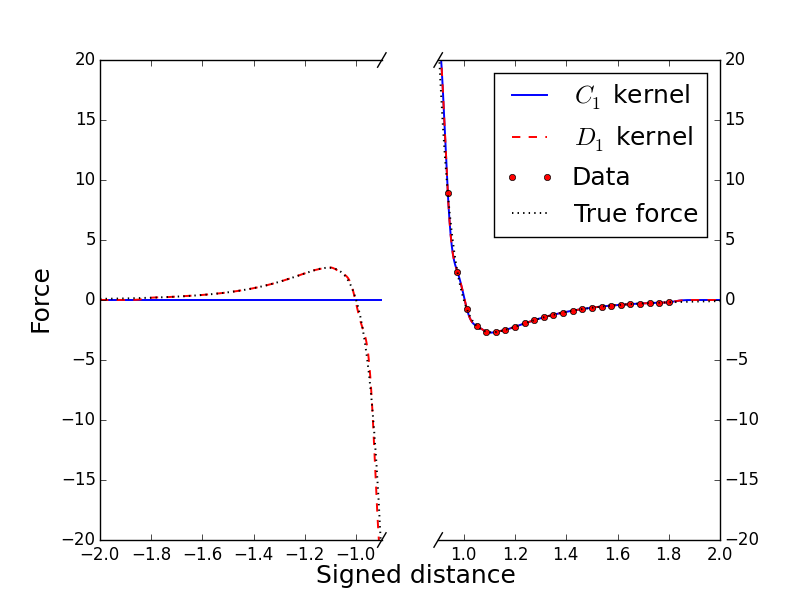}
\par\end{centering}
\caption{Lennard-Jones dimer force field, learned with data from one atom only.
The base kernel ($C_{1}$) does not learn the symmetric counterpart
(reaction force), while the covariant ($D_{1}$) does. The kernels
are labelled by the symmetry group used to make them covariant; see
main text for details. \label{fig: Dimer-force-field.}}
\end{figure}

A key feature of covariant kernels is the ability to enable ``learning''
of the entire set of configurations that are equivalent by symmetry
to those actually provided in the database. For instance, the force
acting on the (``central'') atom at the origin of configuration
$\rho$ can be predicted even if only configurations $\rho'$ of different
symmetry are contained in the database. The only relevant symmetry
transformation in 1D is the reflection $\mathcal{Q}$ of a configuration about
its centre. In the simplest possible system, a dimer, this maps configurations
where the central atom has a right neighbour (i.e.\ those for which
the central atom is the left atom in the dimer) onto configurations
where the central atom has a left neighbour. The covariant symmetrisation
discussed in the previous section [Eq. (\ref{eq: covariant summation})]
takes the very simple form
\begin{equation}
k^{c}(\rho,\rho')=\frac{1}{2}[k_{L}^{b}(\rho,\rho')-k_{L}^{b}(\rho,\mathcal{Q}\rho')].\label{eq: cov ker 1D}
\end{equation}
Note that $k^{c}$ is identically zero for inversion-symmetric
configurations $\rho$ or $\rho'$ whose associated forces must vanish.

The force field associated with a 1D Lennard Jones dimer is plotted
in Fig. \ref{fig: Dimer-force-field.} (dotted curve) as a function
of a single signed number -- the 1D vector going from the central atom
to its neighbour. The figure also shows the predictions of the unsymmetrised
base kernel using training data coming from configurations centred
on the left atom only (solid blue curve). This closely reproduces
the true LJ forces in the region where the data are available, and
predicts the pure prior mean (i.e.\ zero) in the symmetry related region,
i.e.\ the left half of the figure. Meanwhile, because of the covariant
constraint (prior information) the GP based on the covariant kernel
learns the left part of the field by just reflecting the right part
appropriately. 
\begin{figure}
\begin{centering}
\includegraphics[width=0.8\columnwidth]{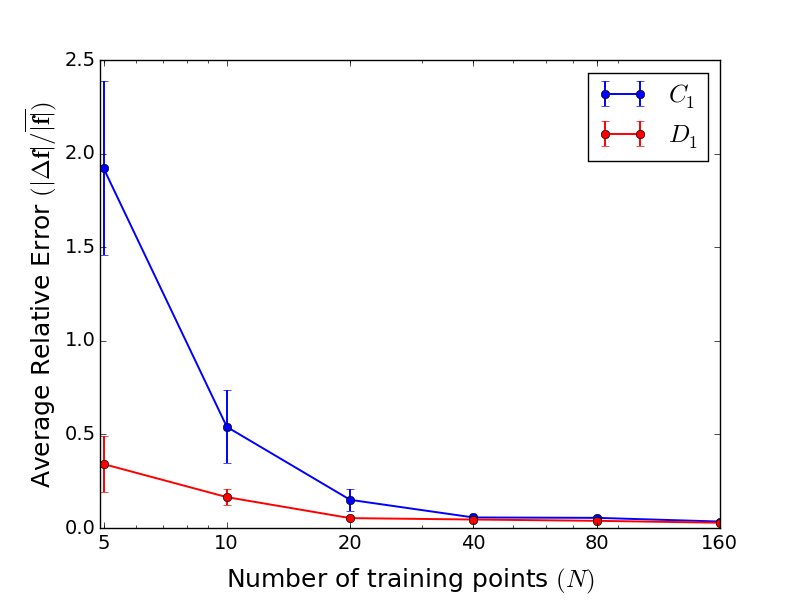}
\par\end{centering}
\caption{Learning Curves for a 1D chain of LJ atoms. The covariant kernel ($D_{1}$)
learns twice as fast as the base one ($C_{1}$). \label{fig :Learning-Curve 1D}}
\end{figure}

To further check the performance of the covariant kernel (\ref{eq: cov ker 1D})
we extended the comparison above to predicting the forces associated
with a 1D Lennard Jones 50-atom chain system, in periodic boundary
conditions. A database of training configurations and an independent
test set of local configurations and forces were sampled from a constant
temperature molecular dynamics simulation using a Langevin thermostat. 

Before presenting the results, it is necessary to introduce some
conventions that will apply throughout the rest of this work. As a
measure of error between reference force $\mathbf{f}^{r}(\rho)$ and
predicted force $\hat{\mathbf{f}}(\rho)$, we will take the absolute
value of their vector difference $|\Delta\mathbf{f}|=|\mathbf{f}^{r}(\rho)-\hat{\mathbf{f}}(\rho)|$.
Relative errors are obtained by dividing this absolute error by the
time-ensemble average of the force modulus $\bar{|\mathbf{f}|}$.
Average errors are found by randomly sampling $N$ training configurations
and $1000$ test configurations. Repeating this operation provides
the standard deviation and hence the error bars on absolute and relative
errors. We furthermore denote by $C_{n}$ the cyclic group of order
$n$ and by $D_{n}$ the dihedral group (containing also reflections)
of order $2n$ ($C_{1}$ hence indicates the trivial group).

With the above clarifications, we can proceed with the analysis of
Figure \ref{fig :Learning-Curve 1D}, which reports the average relative
force error made by the GP regression on the test set as a function of
training set size. It is immediately apparent that the covariant kernel
performance is comparable to that of the base kernel with double the
number of data points for training. We will observe the same effect
also in two and three dimensions: symmetrising over a relevant finite group
of order $|G|$ gives rise to an error drop approximately equivalent
to a $|G|$-fold increase in the number of training points. Since
the computational complexity of training a GP is $\mathcal{O}(N^{3})$,
this can obviously lead to significant computer time savings. 

\subsection{2D systems}

\begin{figure}
\begin{centering}
\includegraphics[width=0.8\columnwidth]{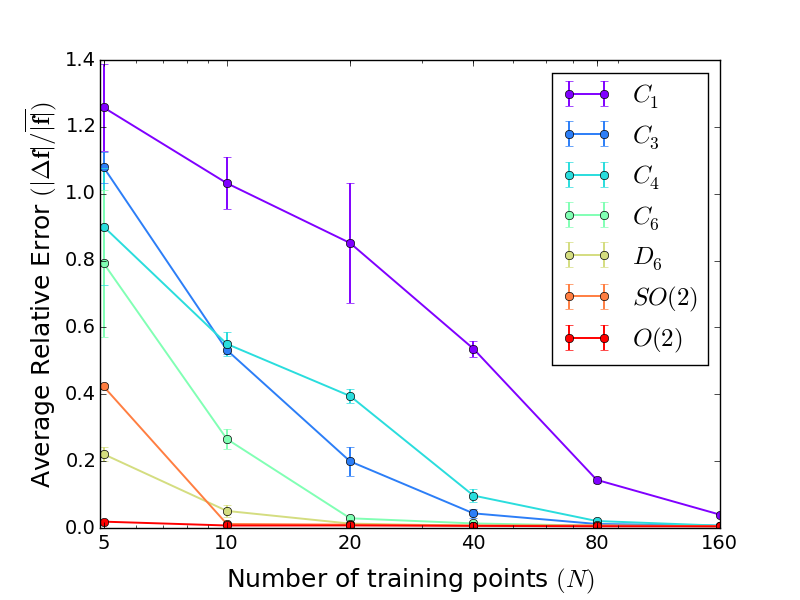}
\par\end{centering}
\caption{Learning curves for 2D triangular grid of LJ atoms. The larger the
symmetry group used to construct the kernel, the faster the learning,
provided that the lattice symmetry is captured. \label{fig: 2D LC}}
\end{figure}

In two dimensions all rotations and reflections, as well as any combination
of these, are elements of $O(2)$. 
Moreover, the $O(2)$ group can be represented by the following set of matrices $O(2)=\{\mathbf{R}(\theta),\,\theta\in(0,2\pi]\}\cup\{\mathbf{R}(\theta)\mathbf{Q},\,\theta\in(0,2\pi]\} $ where $\mathbf{R}(\theta)=\begin{pmatrix}\cos(\theta) & \sin(\theta)\\
-\sin(\theta) & \cos(\theta)
\end{pmatrix}$ and $\mathbf{Q}$ is any $2\times2$ reflection matrix.

This makes the covariant integration (\ref{eq: covariant integration})
over $O(2)$ trivial once the matrix elements resulting from the 
integration over $SO(2)$ have been calculated\@. We next carry out
the integration for the linear base kernel of Eq.~(\ref{eq: base overalp kernel}).
This can be expressed as a sum of pair contributions, where the first
atom in each pair belongs to $\rho$ and the second to $\rho'$ : 
\begin{equation}
\mathbf{K}_{SO(2)}^{c}(\rho,\rho')=\frac{1}{L}\sum_{ij}^{nn'}\int_{SO(2)}d\mathcal{R}\,\mathbf{R}\,\mathrm{e}^{-\frac{\|\mathbf{r}_{i}-\mathbf{R}\mathbf{r}'_{j}\|^{2}}{4\sigma^{2}}}.\label{eq: covint 2d}
\end{equation}
Consistent with Eq.~(\ref{eq: covariant integration}), only one atom
of the pair is rotated during the integration, with $L$ being the
normalisation factor [cf.\ Eq. (\ref{eq: base overalp kernel})].
The pairwise integrals in (\ref{eq: covint 2d}) are calculated in
two steps. We first define $\mathbf{R}_{ij}$ to be the rotation matrix
which aligns $\mathbf{r}_{j}'$ onto $\mathbf{r}_{i}$, and then
perform the change of variable $\tilde{\mathbf{R}}=\mathbf{R}\mathbf{R}_{ij}^{{\rm T}}$
(and analogously $\mathcal{\tilde{R}=\mathcal{R}\mathcal{R}}_{ij}^{-1}$)
yielding 
\begin{equation}
\mathbf{K}_{SO(2)}^{c}(\rho,\rho')=\frac{1}{L}\sum_{ij}\left(\int_{SO(2)}d\mathcal{\tilde{R}}\,\tilde{\mathbf{R}}\,\mathrm{e}^{-\frac{\|\mathbf{r}_{i}-\tilde{\mathbf{R}}\mathbf{R}_{ij}\mathbf{r}'_{j}\|^{2}}{4\sigma^{2}}}\right)\mathbf{R}_{ij}.\label{eq: covint 2d-2}
\end{equation}
Since the two vectors $\mathbf{r}_{i}$ and $\mathbf{R}_{ij}\mathbf{r}_{j}'$
are now aligned, the integral in Eq.~(\ref{eq: covint 2d-2}) can
only depend on the two moduli $r_{i}$ and $r_{j}'$ . The final result
takes a very simple analytic form (cf.\ Supplemental Material): 
\begin{equation}
\mathbf{K}_{SO(2)}^{c}(\rho,\rho')=\frac{1}{L}\sum_{ij}\mathrm{e}^{-\frac{r_{i}^{2}+r_{j}^{\prime2}}{4\sigma^{2}}}I_{1}\left(\frac{r_{i}r_{j}'}{2\sigma^{2}}\right)\mathbf{R}_{ij}\label{eq: SO(2) covariant kernel}
\end{equation}
where $I_{1}(\cdot)$ is a modified Bessel function of the first kind.
The kernel in (\ref{eq: SO(2) covariant kernel}) is rotation-covariant
by construction as can be seen immediately by comparison with Eq.~(\ref{eq: covariant property}). 

By exploiting the internal structure of the orthogonal group discussed above, it is straightforward to show that the roto-reflection covariant kernel is given by
\begin{equation}
\mathbf{K}_{O(2)}^{c}(\rho,\rho')=\frac{1}{2}\left(\mathbf{K}_{SO(2)}^{c}(\rho,\rho')+\mathbf{K}_{SO(2)}^{c}(\rho,\mathcal{Q}\rho')\mathbf{Q}\right),\label{eq: O2 covariant kernel}
\end{equation}
which is the two-dimensional analog of Eq.~(\ref{eq: cov ker 1D}).
Interestingly, the resulting kernel can be also cast in the more intuitive
form 
\begin{equation}
\mathbf{K}_{O(2)}^{c}(\rho,\rho')=\frac{1}{L}\sum_{ij}\mathrm{e}^{-\frac{r_{i}^{2}+r_{j}^{\prime2}}{4\sigma^{2}}}I_{1}\left(\frac{r_{i}r_{j}'}{2\sigma^{2}}\right)\hat{\mathbf{r}}_{i}\hat{\mathbf{r}}_{j}^{\prime{\rm T}},\label{eq: 2D pairwise ker}
\end{equation}
where the hat denotes a normalised vector. 
Equation (\ref{eq: 2D pairwise ker}) implies that the predicted force on an atom at the centre of a configuration $\rho$ will be a sum of pairwise
forces oriented along the directions $\hat{\mathbf{r}}_i$
connecting the central atom with each of its neighbours (while each neighbour will experience a corresponding reaction force). 
The modulus of these forces will be a function of the interatomic
distance completely determined by the training database, whose
integral can be thought of as a pairwise energy potential.
Clearly then, the resulting force field
will be \emph{conservative}: for any fixed
database, the forces predicted by GP inference using this kernel 
will do zero work if integrated along any closed trajectory loop in
configuration space.

To test the relative performance of the learning models discussed
above, we constructed training and test databases for a two-dimensional
triangular lattice, sampled from a constant temperature molecular
dynamics simulation of a 48-particle system interacting via standard
Lennard-Jones forces, once more using periodic boundary conditions
and a Langevin thermostat. As the chosen lattice has three-fold and
six-fold symmetry, we can also examine the performance of covariant
kernels that obey the two properties described above restricted to appropriate finite
groups; these kernels are constructed as in Eq.~(\ref{eq: covariant summation}).
In this way we can monitor how imposing a progressively higher degree
of symmetry on the kernel changes the rate at which forces in this
system can be learned.

Our results are reported in Fig. \ref{fig: 2D LC}. As anticipated,
we find that the discrete covariant summation over the elements of
a group $G$ is approximately equivalent to a $|G|$-fold increase
in the number of data points. This can be seen e.g.\ from the results
for the $C_{3}$ kernel (3-fold rotations) and the $C_{6}$ kernel
(6-fold rotations), by comparing the error incurred in the two cases
using $20$ and $10$ datapoints, respectively. More generally, we
observe that the larger the group, the faster the learning. Note,
however, that for the covariant summation (\ref{eq: covariant summation})
to extract content from the database that is actually useful for predicting
forces in the test configurations at hand, the group used must describe
a true underlying point symmetry of the system. Hence, for instance,
the $C_{4}$ kernel gives rise to much slower learning than the $C_{3}$
kernel for the 2D triangular lattice examined. Consistently, for this
lattice the full point group $D_{6}$ performs almost as well as the
continuous symmetry kernels, suggesting that not much more is to be
gained once the full (finite-group) symmetry
of a system has been captured. This finding enables accurate force
prediction in crystalline system when base kernels are used for which
the covariant integration cannot be performed analytically, because
the summation over a discrete symmetry group is available as a viable
alternative. 

\subsection{3D systems}

\begin{figure}
\begin{centering}
\includegraphics[width=0.8\columnwidth]{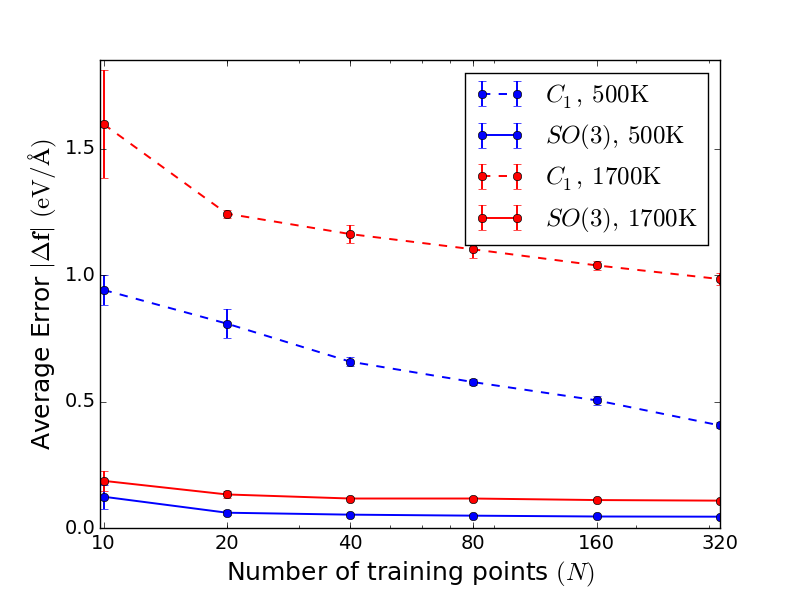}
\par\end{centering}
\caption{Learning Curves for crystalline nickel at two target temperatures.
The $SO(3)$ covariant kernel (full lines) outperforms the base one
(dashed lines). \label{fig: Nickel LC}}
 
\end{figure}

We next benchmark the accuracy of our kernels in predicting DFT forces
in three-dimensional bulk metal systems. As in the 2D case, starting
from the linear base kernel we proceed to carry out the covariant
integration analytically. After expressing the integration as a sum
of pairwise integrals, the position vectors $\mathbf{r}_{i}$ and
$\mathbf{\mathbf{r}}_{j}'$ of two atoms in each pair are aligned
onto each other. A convenient way to achieve this is by making both
vectors parallel to the $z$-axis with appropriate rotations $\mathbf{R}_{i}^{z}$
and $\mathbf{R}_{j}^{z}$. As before, the covariant integration will
yield a matrix whose elements are scalar functions of the radii $r_{i}$
and $r_{j}'$ only. The integration can be carried out analytically
over the standard three Euler angle variables (cf.\ Supplemental Material
for further details). Due to the $z$-axis orientation, the kernel matrix
elements
turn out to be all zero except for the $zz$ one. The result reads
\begin{eqnarray}
\mathbf{K}_{SO(3)}^{c}(\rho,\rho') & = & \frac{1}{L}\sum_{ij}\mathbf{R}_{i}^{z{\rm T}}\begin{pmatrix}0 & 0 & 0\\
0 & 0 & 0\\
0 & 0 & \phi(r_{i},r_{j}')
\end{pmatrix}\mathbf{R}_{j}^{z},\nonumber \\
\phi(r_{i},r_{j}) & = & \frac{\mathrm{e}^{-\alpha_{ij}}}{\gamma_{ij}^{2}}(\gamma_{ij}\cosh\gamma_{ij}-\sinh\gamma_{ij}),\nonumber \\
\alpha_{ij} & = & \frac{r_{i}^{2}+r_{j}^{\prime2}}{4\sigma^{2}},\nonumber \\
\gamma_{ij} & = & \frac{r_{i}r_{j}'}{2\sigma^{2}}. \label{eq: SO3 covariant kernel}
\end{eqnarray}
As in the 2D case, this covariant kernel matrix can be rewritten
in terms of the unit vectors $\hat{\mathbf{r}}_{i}$ and $\hat{\mathbf{r}}_{j}^{\prime}$
associated with the atoms of the configurations $\rho,\rho'$ as
\begin{equation}
\mathbf{K}_{SO(3)}^{c}(\rho,\rho')=\frac{1}{L}\sum_{ij}\phi(r_{i},r_{j}^{\prime})\hat{\mathbf{r}}_{i}\hat{\mathbf{r}}_{j}^{\prime{\rm T}},\label{eq: 3D pairwise ker}
\end{equation}
making it apparent that the kernel models a pairwise conservative
force field. However, while in 2D we needed to impose the full roto-reflection
symmetry in order to obtain Eq.~(\ref{eq: 2D pairwise ker}),
rotations alone are sufficient to arrive at the fully covariant kernel
in (\ref{eq: 3D pairwise ker}). This is a consequence of the fact
that, in three dimensions, the covariant integral over rotations already imposes
that the predicted force any atom will exert on any other is aligned
along the vector connecting the pair: by symmetry there can be no
preferred direction for an orthogonal force component after integrating
over all rotations around the connecting vector, so that $\mathbf{K}_{O(3)}^{c}=\mathbf{K}_{SO(3)}^{c}$.
This is not the case in two dimensions where covariant integration is over rotations
around the $z$-axis orthogonal to all connecting vectors lying in
the $xy$ plane, so that non-aligned predicted force components associated
with a non-zero torque are not forbidden by symmetry in $\mathbf{K}_{SO(2)}^{c}$,
and only the fully symmetrised kernel (\ref{eq: O2 covariant kernel})
will reduce to the pairwise form (\ref{eq: 2D pairwise ker}). 
\begin{figure}
\begin{centering}
\includegraphics[width=0.8\columnwidth]{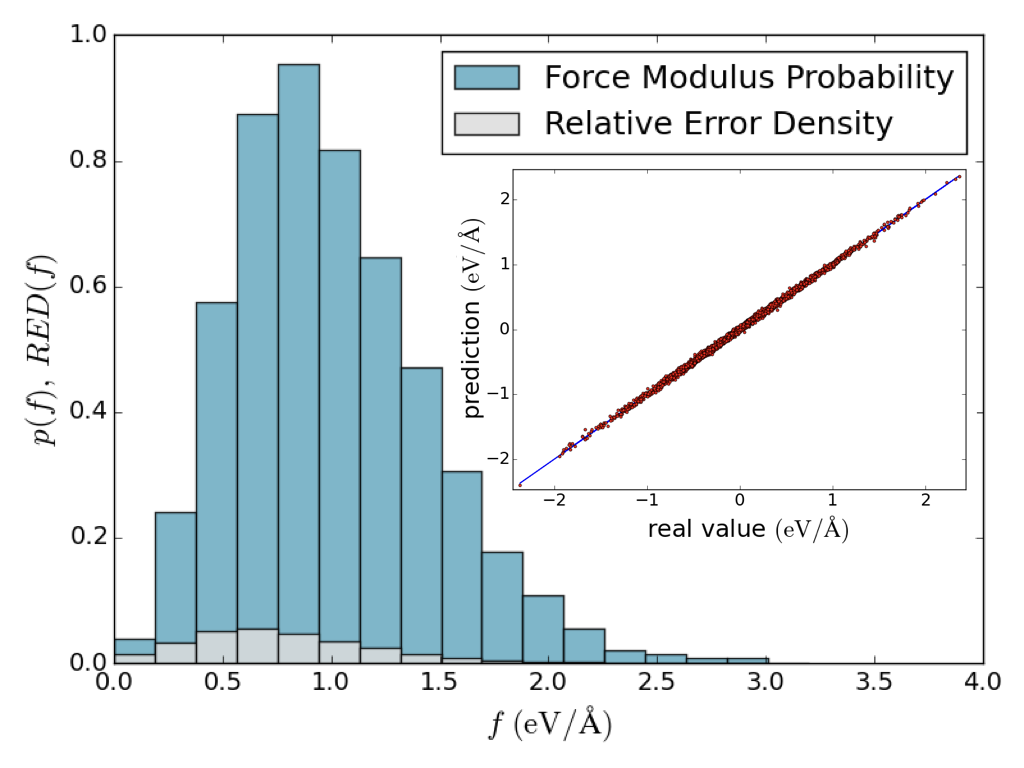}
\par\end{centering}
\caption{Density of relative error made by the GP algorithm ($N=320$) for
bulk nickel at $500 \rm{K}$. The inset shows the scatter plot of real vs.\
predicted cartesian components for the same data. \label{fig: Nickel DRE}}
\end{figure}
More generally we may conjecture that the rotationally covariant kernel
$\mathbf{K}_{O(d)}^{c}$ derived from a linear base kernel
predicts pairwise central forces, and hence is conservative, in any
dimension $d$. 

We note that energy conserving kernels have previously been 
obtained as double derivatives (Hessian matrices) of scalar energy
kernels (as originally described in Refs. 
  \cite{Narcowich:2007tl,Macedo:2008vq}
and used for atomistic systems in Refs. 
in \cite{Bartok:2015iw} to learn energies and more recently 
in Ref. \cite{Chmiela:2017ff}  to learn forces).
However, no closed-form expressions exist for the energy kernels that would yield our $O(d)$ energy conserving kernels through this route, since the required double integration of the kernels (\ref{eq: cov ker 1D}), (\ref{eq: 2D pairwise ker}), (\ref{eq: 3D pairwise ker}) cannot be carried out analytically.

To test our models, we performed DFT-accurate dynamical simulation
with exchange and correlation energy modeled via the PBE/GGA approximation \cite{Perdew:1996iq}.
The systems considered were $4\times4\times4$ supercells of fcc nickel
and bcc iron in periodic boundary conditions.
A weakly coupled Langevin thermostat was used to control the temperature.  
We first examine bulk nickel at the target temperatures of $500{\rm K}$
and $1700{\rm K}$, i.e.\ for an intermediate temperature where anharmonic
behaviour is already significant, and at a temperature close to the
melting point where the strong thermal fluctuations make the system 
explore a more complex target configuration space. 
Figure \ref{fig: Nickel LC} illustrates the performance
of the kernel in Eq.~(\ref{eq: SO3 covariant kernel}) on this system. 

The effect of adding symmetry information on the learning curve is
very significant for both temperatures. In particular, the $SO(3)$
covariant kernel achieves a force error average lower than the $0.1\rm{eV}/\text{\AA}$
threshold using remarkably few training points: $10$ and $80$ for
the lower and higher temperatures in this test, respectively. The
errors of the most accurate models (achieved with a $N=320$ database)
are particularly low: $0.0435(\pm 0.0006)\rm{eV}/\text{\AA}$ and
$0.095(\pm0.003)\rm{eV}/\text{\AA}$ respectively. Moreover, we note
that the error on each force component (often reported in the literature,
and different from the error on the full force vector used here) will
be lower by a factor $\sqrt{3}$. This yields errors of
$0.025\rm{eV}/\text{\AA}$ and $0.052\rm{eV}/\text{\AA}$ in the two
cases, the former comparing well with the
 $0.09\rm{eV}/\text{\AA}$ value obtained by using a state of the 
art Embedded Atom Model (EAM) interatomic potential for
nickel \cite{Mishin:2004bh, Bianchini:2016jn}.

Figure \ref{fig: Nickel DRE} allows one to assess the accuracy of
the GP predictions in a complementary way: here we plot the probability
distribution of the atomic forces as a function of the force modulus
(blue histogram) and the associated relative error density (grey histogram).
We define the latter as $RED(f)=\frac{|\Delta\mathbf{f}|}{f}p(f)$,
which is normalised to $0.055$, reflecting the $5.5\%$ average relative
error incurred by force prediction. The fact that $RED(f)$ is everywhere
a small fraction of $p(f)$ demonstrates that a reasonable accuracy
is achieved for the whole range of forces predicted. 

The results presented so far indicate that fully exploiting
symmetry significantly improves the accuracy of force prediction. 
Covariance is thus always used in the following analysis, where we
compare the performance of different symmetry-aware kernels.
We start by choosing iron systems for
these tests as many properties of iron-based systems remain out of
modelling reach. This is mostly due to technical limitations. On the one hand, full DFT calculations on large systems are too computationally expensive and even hybrid quantum-classical (``QM/MM'')
simulations of iron systems are typically overwhelmingly costly, as
they require large QM-zone buffered clusters to fully converge the
forces \cite{Bianchini:cxBexsLf}. On the other hand, in many
situations even the best available, state of the art classical force
fields may not guarantee accurate force prediction, as they may incur
systematic errors \cite{Bianchini:cxBexsLf,Bianchini:2016jn}, or
may be hard to extend to complex chemical compositions \cite{vonPezold:2011iz},
so that a technique that can indefinitely re-use all computed QM forces
via GP inference and produce results that are traceably aligned with
DFT-accurate forces could be very useful \cite{Szlachta:2014jh,Li:2015eb}. 

\begin{figure}
\begin{centering}
\includegraphics[width=0.8\columnwidth]{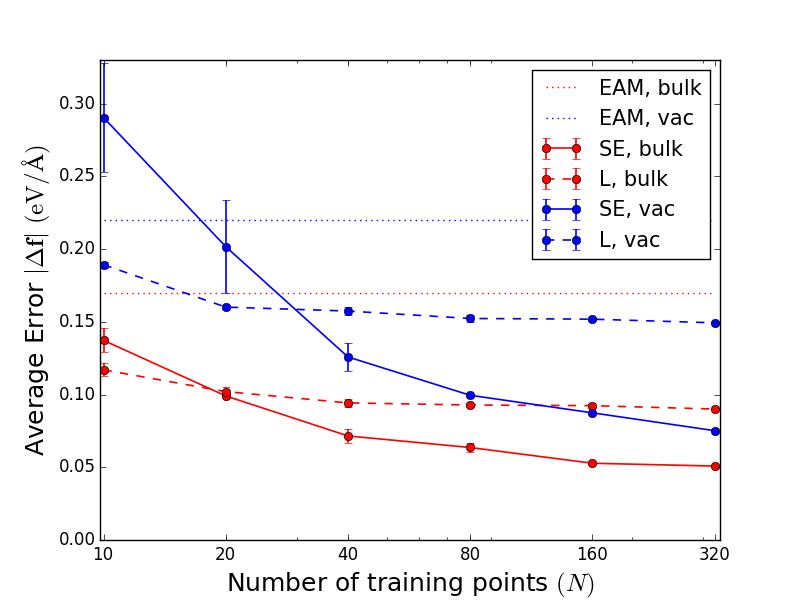}
\par\end{centering}
\caption{Learning curves associated with force prediction by the linear (L, dashed lines)
and squared exponential (SE, solid lines) covariant kernels in bulk iron systems.
Red and blue colours indicate undefected systems and model systems
containing a vacancy, respectively. \label{fig: Iron LC}}
\end{figure}

We carried out constant temperature $(500{\rm K}$) molecular dynamics
simulations of two bcc iron systems: a 64-atom crystalline system
and a 63-atom system derived from this and containing a single
vacancy. 
In the latter, only the atoms within the first two neighbour
shells of the vacancy were used to test the algorithm, to better resolve 
the performance of our kernels in a defective system. 
Figure \ref{fig: Iron LC} shows the learning curves for the two symmetrised kernels: the linear
kernel covariant over $O(3)$ and the squared exponential kernel (\ref{eq: SE kernel}) covariant over the full cubic point-group of the crystal. The figure
also reports the performance of a high-quality
EAM potential \cite{Mendelev:2003co}. Both kernels perform better
than the EAM potentials in this test. However, the error rate of the
linear kernel (dashed lines) levels off to some constant non-zero
value that might or might not be satisfactory (depending on the application),
and will generally depend on the system being examined. In bulk iron
the error floor value is about $0.09{\rm eV}/\text{\AA}$ while in
the vicinity of a vacancy it is considerably higher ($0.15{\rm eV}/\text{\AA}$),
suggesting that in spite of its many attractive properties (e.g.\ fast
evaluation, fast convergence, energy conservation), the linear class
of kernels of the form (\ref{eq: 3D pairwise ker}) is by no means
complete, that is, it sometimes cannot capture and reproduce the entirety
of the reference QM physical interaction. In many situations, kernels
capable of reproducing higher order interactions could be needed to
reach the target accuracy. This is exemplified by the much better
performance of the squared exponential kernel (full lines in the figure),
which yields higher accuracy, particularly for the more complex vacancy
system (about $0.05{\rm eV}/\text{\AA}$ and $0.075{\rm eV}/\text{\AA}$
for atoms in the bulk and near the vacancy respectively). 
It is worth noting here that, in general, conserving energy exactly by construction 
provides no guarantee of higher force accuracy.  
For instance, in the case above, the squared exponential kernel delivers much more precise forces even though it conserves energy only approximately. 
As the approximation will in any case improve with the accuracy of the predicted forces, 
while no $SO(3)$-invariant energy conserving equivalent of this kernel
has been proposed or appears viable, whether it is preferable
to use this kernel or a less accurate but energy
conserving alternative one, will generally depend on both the target system and
the application at hand.

\begin{figure}
	\begin{centering}
		\includegraphics[width=0.8\columnwidth]{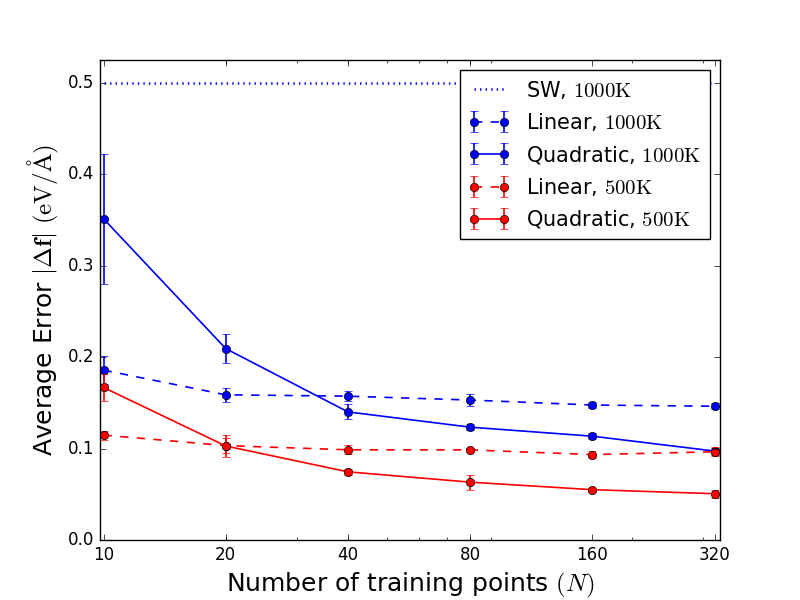}
		\par\end{centering}
	\caption{Learning curves obtained 
		for crystalline silicon using the linear kernel (dashed lines) or the quadratic kernel (solid lines). Different colours indicate different temperatures. \label{fig: Silicon_LC}}
\end{figure}

For target systems with no clear point symmetry, a full covariant
integration would always be desirable. 
This cannot be carried out analytically for the squared exponential
kernel, where symmetrising by a discrete summation is the only 
option.
However, interactions beyond pairwise can be still captured by 
the \emph{quadratic} kernel obtained by taking the square of the 
linear kernel (\ref{eq: base overalp kernel}). 
In contrast to the squared exponential kernel, this is analytically 
tractable (for instance, an $SO(3)$-invariant scalar quadratic 
kernel was obtained in \cite{,Bartok:2013cs}), and our analysis 
reveals that a matrix-valued quadratic kernel covariant over $O(3)$ can
be derived analytically (details of the calculation are a subject for future work  \cite{Glielmo:kZKvR16V}).
The resulting model generates a roto-reflection symmetric three-body
force field that can be expected to properly describe non close-packed
bonding, such as found in covalent systems, for example.

Figure \ref{fig: Silicon_LC} illustrates the errors incurred 
by the linear and the quadratic kernel while attempting to reproduce 
the forces obtained during Langevin dynamics of a 64-atom 
crystalline silicon system using Density Functional Tight Binding (DFTB)
\cite{Elstner:1998gh}.  
Both linear and quadratic kernel are significantly more accurate 
than a classical Stillinger Weber (SW) potential
\cite{,Stillinger:1985zz} fitted to reproduce the 
DFTB lattice parameter and bulk modulus \cite{Li:2015eb}. 
Due to its more restricted associated function space, the linear
kernel is the one that learns faster, and would be the more accurate 
if only very restricted databases had to be used. 
However, the quadratic kernel eventually performs much 
better than the (effective two-body) linear one for both 
of the temperatures, $500\rm{K}$ and
$1000\rm{K}$, that we investigated in this covalent system. We obtain
errors 
of $0.05{\rm eV}/\text{\AA}$ and $0.1{\rm eV}/\text{\AA}$
in the two cases, corresponding respectively to approximately $4\%$ and $6\%$ 
of the mean force.  
These are very close to the minimum baseline \emph{locality error}
\cite{Deringer:2017ea} associated with the finite cutoff radius used for the 
Gaussian expansion in (\ref{eq: configuration expansion}).

\section{Conclusion}

In this work we presented a new method to learn quantum forces on
local configurations. 
This method is based on a Vectorial Gaussian
Process that encodes prior knowledge in a matrix valued kernel function.
We showed how to include rotation and reflection symmetry of the force
in the GP process via the notion and use of covariant kernels. 
A general recipe was provided to impose this property on otherwise non-symmetric
kernels. 
The essence of this recipe lies in a special integration
step, which we call covariant integration, over the full roto-reflection
group associated with the relevant number of system dimensions. 
This calculation can be performed analytically starting from a linear base
kernel, and the resulting $O(d)$ covariant kernels can be shown
to generate conservative force fields.

We furthermore tested covariant kernels on standard physical systems
in one, two and three dimensions. 
The one- and two-dimensional scenarios served as playgrounds to better understand and illustrate the essential features of such learning. 
The 3D systems allowed some practical benchmarking
of the methodology in real systems. 
In agreement with what physical
intuition would suggest, we consistently found that incorporating
symmetry gives rise to more efficient learning. 
In particular, if
both database and target configurations belong to a system with 
a definite underlying symmetry, restricting kernel covariance to the 
corresponding finite symmetry group will deliver the full speed-up of error convergence
with respect to database size. At the same time this approach
lifts the requirement of 
analytical integrability over the full $SO(d)$ manifold, as the
restricted integration becomes a simple discrete sum over the relevant finite
set of group elements. 
Testing on nickel, silicon and iron (the latter both pure and defective) reveals that the present recipes can improve significantly on available classical potentials.
In general, non-linear kernels may be needed for
accurate force predictions in the presence of complicated
interactions, e.g.\ in the study of plasticity or
embrittlement/fracture behaviour of covalent or metallic systems. 
In particular, a quadratic base kernel yields 
a fully $O(3)$ covariant effective three-body force field, and our
tests suggest that this can be used successfully to improve the accuracy 
of force prediction in covalent materials. 
Current work is focussing on amorphous Si systems, where the lack of a clear point symmetry 
makes the full $O(3)$ covariance strictly necessary.

Our results reveal that force covariance is achievable without imposing energy conservation to the kernel form.  
While both are desirable properties, we find that lifting the exact energy conservation constraint can sometimes yield higher force accuracy.
For instance, no invariant local energy based kernel has been proposed for the squared exponential ("universal approximator") kernel, since the analytic integration over $SO(3)$ is not viable. 
However, we find that covariance limited to the $O_{48}$ point group is very effective for force predictions in crystalline Fe systems using this kernel (see Fig. \ref{fig: Iron LC}).

In general, while predicting forces with high accuracy is the main motivation for machine learning-based work in this field, the best compromise between accuracy, energy conservation and covariance will depend on the specific target application.
For instance, kernels built from a covariant integration (or summation) that do not conserve energy exactly should not be used as substitutes for conventional interatomic potentials to perform long NVE simulations, since they might in principle lead to non-negligible spurious energy drifts.  
This is not a problem in NVT simulations, where a thermostat exchanges energy with the system to achieve and conserve the target temperature, which will be able to compensate for any such drift if appropriately chosen \cite{Jones:2011fh}.  
Furthermore, the same kernels will be particularly suited for 
schemes that are in all cases incompatible with strict energy conservation.  
These include the LOTF approach and any online learning scheme similarly involving a dynamically updated force model.
They also include any highly accurate and transferable scheme based on a fixed, very large database where, to maximise efficiency, each force prediction only uses its corresponding most relevant  database subset.

On the other hand, any usage style is possible for covariant kernels conserving energy exactly, such as the covariant linear kernels of Eqs. (\ref{eq: cov ker 1D}), (\ref{eq: 2D pairwise ker}), and (\ref{eq: 3D pairwise ker}).   
In fact, the conservative pairwise interaction forces generated by these covariant linear kernels can be easily integrated to provide effective ``optimal'' standard pairwise potentials for any application needing a total energy expression.
We also note that while the pair interaction form would still ensure very fast evaluation of the predicted forces, its accuracy for complex systems could be improved by dropping the transferability requirement of a single pairwise function.  
    In such a scheme, different
    system regions could conceivably be modelled by locally optimised forces/potentials, where the local tuning could be simply achieved by restricting the inference process to subsets of the database pertinent to each target region.

\section*{Acknowledgements}

The authors acknowledge funding by the Engineering and Physical Sciences
Research Council (EPSRC) through the Centre for Doctoral Training ``Cross Disciplinary Approaches to Non-Equilibrium Systems'' (CANES,
Grant No. EP/L015854/1), by the Office of Naval Research Global (ONRG
Award No. N62909-15-1-N079). ADV acknowledges further support by the
EPSRC HEmS Grant No. EP/L014742/1 and by the European Union\textquoteright s
Horizon 2020 research and innovation program (Grant No. 676580, The
NOMAD Laboratory, a European Centre of Excellence). The research used
resources of the Argonne Leadership Computing Facility at Argonne
National Laboratory, which is supported by the Office of Science of
the U.S. Department of Energy under Contract No. DE-AC02-06CH11357.
The DFT dataset used in this work is openly available from the research data management system of King's College London at http://doi.org/10.18742/RDM01-92.
AG would like to thank F. Bianchini for his help in data collection
and R. G. Margiotta, K. Rossi, F. Bianchini and C. Zeni for useful
discussions.

\bibliographystyle{IEEEtran}
\bibliography{Covariant2.bib}

\newpage

\section*{Supplemental Material}

\subsection*{Covariant integration}

The integral we wish to evaluate, repeated here for convenience, is

\begin{eqnarray*}
	\mathbf{K}^{c}\ (\rho,\rho') & = & \frac{1}{L}\sum_{ij}\int d\mathcal{R}\,\mathbf{R}\mathrm{e}^{-(\mathbf{r}_{i}-\mathbf{Rr}_{j}')^{2}/4\sigma^{2}}\\
	& = & \frac{1}{L}\sum_{ij}I_{ij}.
\end{eqnarray*}
First of all it is convenient to separate the radial part from the
angular one as the first of these does not depend on rotations:
\begin{eqnarray*}
	I_{ij} & = & \mathrm{e}^{-(r_{i}^{2}+r_{j}^{\prime2})/4\sigma^{2}}\int d\mathcal{R}\,\mathbf{R}\,\mathrm{e}^{\mathbf{r}_{i}^{{\rm T}}\mathbf{Rr}_{j}'/2\sigma^{2}}\\
	& = & C_{ij}\int d\mathcal{R}\,\mathbf{R}\,\mathrm{e}^{\mathbf{r}_{i}^{{\rm T}}\mathbf{Rr}_{j}'/2\sigma^{2}}.
\end{eqnarray*}

\subsubsection*{2D systems}

If we define $\mathbf{R}_{ij}$ to be the rotation matrix that brings
the vector $\mathbf{r}_{j}'$ onto $\mathbf{r}_{i}$, then we can
perform the change of variable $\tilde{\mathbf{R}}=\mathbf{R}\mathbf{R}_{ij}^{{\rm T}}$
\begin{eqnarray*}
	I_{ij} & = & C_{ij}\int d\mathcal{\tilde{R}}\,\tilde{\mathbf{R}}\,\mathrm{e}^{\mathbf{r}_{i}^{{\rm T}}\tilde{\mathbf{R}}\mathbf{R}_{ij}\mathbf{r}_{j}'/2\sigma^{2}}\mathbf{R}_{ij}\\
	& = & C_{ij}\int d\mathcal{\tilde{R}}\,\tilde{\mathbf{R}}\,\mathrm{e}^{\mathbf{r}_{i}^{{\rm T}}\tilde{\mathbf{R}}\tilde{\mathbf{r}}_{j}'/2\sigma^{2}}\mathbf{R}_{ij}.
\end{eqnarray*}
where the two vectors $\mathbf{r}_{i}$ and $\tilde{\mathbf{r}}_{j}$
are now aligned with each other. By parametrising all rotations by
a single angle $\theta$ we can rewrite the above integration as
\begin{eqnarray*}
	I_{ij} & = & C_{ij}\int_{0}^{2\pi}\frac{d\theta}{2\pi}\mathbf{\,R}(\theta)\mathrm{e}^{\mathbf{r}_{i}^{{\rm T}}\mathbf{R}(\theta)\tilde{\mathbf{r}}_{j}'/2\sigma^{2}}\mathbf{R}_{ij}\\
	& = & C_{ij}\left(\int_{0}^{2\pi}\frac{d\theta}{2\pi}\mathbf{\,R}(\theta)\mathrm{e}^{r_{i}r_{j}'\cos\theta/2\sigma^{2}}\right)\mathbf{R}_{ij}.
\end{eqnarray*}
The integral in brackets can now be given an analytic form. The rotation
matrix $\mathbf{R}(\theta)$ is composed by $\cos\theta$ on the diagonal
and $\{\sin\theta,-\sin\theta\}$ off the diagonal. Evaluating the
above integration for such terms one finds that 
\[
\begin{cases}
\int_{0}^{2\pi}\frac{d\theta}{2\pi}\cos\theta\,\mathrm{e}^{r_{i}r_{j}'\cos\theta/2\sigma^{2}} & =I_{1}\left(\frac{r_{i}r_{j}'}{2\sigma^{2}}\right)\\
\int_{0}^{2\pi}\frac{d\theta}{2\pi}\sin\theta\,\mathrm{e}^{r_{i}r_{j}'\cos\theta/2\sigma^{2}} & =0
\end{cases}
\]
where $I_{1}(\cdot)$ is a modified Bessel function of the first kind.
The second line follows because we are integrating an odd function
over an even domain. The first line, on the other hand, results from
a definition of modified Bessel functions of the first kind $I_{n}(z)$
for integer values of $n$ (\cite{Abramowitz:1972vl} p. 376), i.e.
\[
I_{n}(z)=\frac{1}{\pi}\int_{0}^{\pi}\mathrm{e}^{z\cos\theta}\cos(n\theta)d\theta.
\]
Hence the final integral reads 
\[
I_{ij}=C_{ij}I_{1}\left(\frac{r_{i}r_{j}'}{2\sigma^{2}}\right)\mathbf{R}_{ij}.
\]

\subsubsection*{3D systems}

In three dimensions, it is first of all convenient to cast the integral
in the following form 
\begin{eqnarray*}
	I_{ij} & = & \int d\mathcal{R}\,\mathbf{R}\mathrm{e}^{-(\mathbf{r}_{i}-\mathbf{Rr}_{j}')^{2}/4\sigma^{2}}\\
	& = & \int d\mathcal{R}\,\mathbf{R}\,k^{p}(\mathbf{r}_{i},\mathbf{Rr}_{j}').
\end{eqnarray*}
Now we can use the \emph{global invariance} of the base pairwise kernels
$k^{p}$, that is $k^{p}(\mathbf{r},\mathbf{r}')=k^{p}(\mathbf{Rr},\mathbf{Rr}')$,
in order to align $\mathbf{r}_{i}$ onto the $z$-axis. We call the
rotation that does so $\mathbf{R}_{i}^{z}$ and we have
\begin{eqnarray*}
	I_{ij} & = & \int d\mathcal{R}\,\mathbf{R}\,k^{p}(\mathbf{R}_{i}^{z}\mathbf{r}_{i},\mathbf{R}_{i}^{z}\mathbf{Rr}_{j}')\\
	& = & \int d\mathcal{R}\,\mathbf{R}\,k^{p}(\tilde{\mathbf{r}}_{i},\mathbf{R}_{i}^{z}\mathbf{R}\mathbf{r}_{j}').
\end{eqnarray*}
where we defined $\tilde{\mathbf{r}}_{i}=\mathbf{R}_{i}^{z}\mathbf{r}_{i}$.
At this point we find the matrix $\mathbf{R}_{j}^{z}$ that brings
also $\mathbf{r}_{j}$ parallel to the $z$-axis. We then insert it
in front of $\mathbf{r}_{j}'$ in the form of the identity $\mathbf{R}_{j}^{z{\rm T}}\mathbf{R}_{j}^{z}$:
\begin{eqnarray*}
	I_{ij} & = & \int d\mathcal{R}\,\mathbf{R}\,k^{p}(\tilde{\mathbf{r}}_{i},\mathbf{R}_{i}^{z}\mathbf{R}\mathbf{R}_{j}^{z{\rm T}}\mathbf{R}_{j}^{z}\mathbf{r}_{j}')\\
	& = & \int d\mathcal{R}\,\mathbf{R}\,k^{p}(\tilde{\mathbf{r}}_{i},\mathbf{R}_{i}^{z}\mathbf{R}\mathbf{R}_{j}^{z{\rm T}}\tilde{\mathbf{r}}_{j}')
\end{eqnarray*}
where we again used the tilde notation to define the vector now aligned
to the $z$-azis. Finally we perform the change of variables $\tilde{\mathbf{R}}=\mathbf{R}_{i}^{z}\mathbf{R}\mathbf{R}_{j}^{z{\rm T}}$
to obtain 
\begin{eqnarray*}
	I_{ij} & = & \mathbf{R}_{i}^{z{\rm T}}\int d\tilde{\mathcal{R}}\,\tilde{\mathbf{R}}\,k^{p}(\tilde{\mathbf{r}}_{i},\tilde{\mathbf{R}}\tilde{\mathbf{r}}_{j}')\mathbf{R}_{j}^{z}\\
	& = & \mathbf{R}_{i}^{z{\rm T}}\mathbf{R}_{ij}\mathbf{R}_{j}^{z}.
\end{eqnarray*}
The central integral yielding $\mathbf{R}_{ij}$ remains to be performed.
Its evaluation is considerably simpler than the original problem since
now both vectors $\tilde{\mathbf{r}}_{i},\tilde{\mathbf{r}}_{j}'$
are along the $z$-axis. Hence, by parametrising all rotations by
Euler angles $\alpha,\beta,\gamma$ around the $z,y,z$ axes respectively,
we find by geometric reasoning that the argument of the exponential
has to be invariant upon rotations of angles $\alpha$ and $\gamma$
around the $z$-axis. In fact, we have that
\[
\mathbf{R}_{ij}=C_{ij}\int\frac{d\alpha d\beta d\gamma\sin\beta}{8\pi^{2}}\mathbf{R}(\alpha,\beta,\gamma)\mathrm{e}^{r_{i}r_{j}'\cos\beta/2\sigma^{2}}
\]
where we made use of the normalised Haar measure $d\alpha d\beta d\gamma\sin\beta/8\pi^{3}$.
The rotation matrix to be averaged reads 
\[
\mathbf{R}(\alpha,\beta,\gamma)=\begin{pmatrix}c_{\alpha}c_{\gamma}-c_{\beta}s_{\alpha}s_{\gamma} & -c_{\gamma}c_{\beta}s_{\alpha}-c_{\alpha}s_{\gamma} & s_{\alpha}s_{\beta}\\
c_{\gamma}s_{\alpha}+c_{\alpha}c_{\beta}s_{\gamma} & c_{\alpha}c_{\gamma}c_{\beta}-s_{\alpha}s_{\gamma} & -c_{\alpha}s_{\beta}\\
s_{\gamma}s_{\beta} & c_{\gamma}s_{\beta} & c_{\beta}
\end{pmatrix}.
\]
All the elements of the above matrix apart from the $zz$ element
vanish since there is always either a sine or a cosine integrated
over an entire period. By defining $\gamma_{ij}=r_{i}r_{j}'/2\sigma^{2}$,
the only non trivial integral reads
\begin{eqnarray*}
	\int_{0}^{\pi}\frac{d\beta\sin\beta}{2}\cos\beta\mathrm{e}^{r_{i}r_{j}'\cos\beta/2\sigma^{2}} & = & \int_{0}^{\pi}\frac{d\beta}{2}\frac{\sin(2\beta)}{2}\mathrm{e}^{\gamma_{ij}\cos\beta}\\
	& = & \left[\frac{\mathrm{e}^{\gamma_{ij}\cos\beta}(1-\gamma_{ij}\cos\beta)}{2\gamma_{ij}^{2}}\right]_{0}^{\pi}\\
	& = & \frac{\gamma_{ij}\cosh\gamma_{ij}-\sinh\gamma_{ij}}{\gamma_{ij}^{2}}.
\end{eqnarray*}

\end{document}